\documentclass[a4paper,12pt]{article}
\usepackage{amsmath,amssymb}
\usepackage[dvips]{graphicx}

\newcommand{\bea}{\begin{eqnarray}}
\newcommand{\ena}{\end{eqnarray}}

\newcommand{\e}{\epsilon}
\def\bbox{{\,\lower0.9pt\vbox{\hrule \hbox{\vrule height 0.2 cm
\hskip 0.2 cm \vrule height 0.2 cm}\hrule}\,}}
\newcommand{\dsl}{\pa \kern-0.5em /}

\newcommand{\pa}{\partial}

\newcommand{\nn}{\nonumber\\}
\newcommand{\p}[1]{(\ref{#1})}

\begin{document}

\topmargin -10mm
\oddsidemargin 0mm

\begin{flushright}
YITP-02-26 \\
OU-HET 409 \\
hep-th/0204161
\end{flushright}

\vspace{1.5cm}
\begin{center}
{\LARGE\bf Supertubes and Supercurves from M-Ribbons} \\
\vspace{1.5cm}
{\large
Yoshifumi Hyakutake\footnote{e-mail address: hyaku@yukawa.kyoto-u.ac.jp}
and
Nobuyoshi Ohta\footnote{e-mail address: ohta@phys.sci.osaka-u.ac.jp}} \\
\vspace{1.5cm}
${}^1${\em Yukawa Institute for Theoretical Physics, Kyoto University} \\
              {\em Sakyo-ku, Kyoto 606-8502, Japan}\\[0.5cm]
${}^2${\em Department of Physics, Osaka University,
              Toyonaka, Osaka 560-0043, Japan} \\
\vspace{1cm}

\large{ABSTRACT}
        \par
\end{center}

\begin{quote}
\begin{normalsize}
We construct $\frac{1}{4}$ BPS configurations, `M-ribbons', in M-theory on
$T^2$, which give the supertubes and supercurves in type IIA theory upon
dimensional reduction. These M-ribbons are generalized so as to be consistent
with the $SL(2,\mathbb{Z})$ modular transformation on $T^2$.
In terms of the type IIB theory, the generalized M-ribbons are interpreted as
an $SL(2,\mathbb{Z})$ duality family of super D-helix.
It is also shown that the BPS M-ribbons must be straight in one direction.

\end{normalsize}
\end{quote}

\newpage

\renewcommand{\thefootnote}{\arabic{footnote}}
\setcounter{footnote}{0}

\section{Introduction}

Recently an interesting class of $\frac{1}{4}$ BPS brane configurations,
supertube, super D-helix and supercurve, have been studied
\cite{MT,BL,CO,EMT,BKi,MNT1,MNT2}. The supertube is a tubular D2-brane with
an arbitrary cross-section in 10-dimensional flat space-time and is supported
against collapse by the existence of both electric and magnetic fluxes on it.
The super D-helix and supercurve are constructed by making T-duality and
T-S-T-duality on the supertube, respectively.\footnote{The original
super D-helix is a helical D-string, but here we include configurations
with an arbitrary curve.}
One of the interesting aspects of the supertube is that it can be deformed
into a pair of D2-anti-D2 system, which still preserves quarter
supersymmetry~\cite{BK}. In fact, open strings ending on both D2 and anti-D2
do not contain tachyonic modes and hence the configuration is stable~\cite{BO}.
It is also possible to extend such supersymmetric D-anti-D system to
higher-dimensional D-branes~\cite{BOS}.

The electric and magnetic fluxes on the D2-brane correspond to the charges of
fundamental strings and D0-branes dissolved into the D2-brane world-volume.
Therefore the supertube is a bound state of D2-brane, fundamental strings
and D0-branes, which is regarded as a variant of M2-brane in the context
of M-theory. In view of the above connections of those various configurations
under duality transformations, it is interesting to give a unified view of
those in the M-theory perspective.
In this paper, we construct classical $\frac{1}{4}$ BPS configurations
of an M2-brane in M-theory compactified on $T^2$, say 9th and 11th directions,
and show that our configurations give the above supertube and supercurve in
type IIA theory upon dimensional reduction.\footnote{Similar method was
applied to obtain classical configurations of M2-brane in the background
of M5-branes in ref.~\cite{Hya1}, some of which are regarded as expanded
strings in the background of NS5-branes in type IIA theory.} We call these
M2-brane configurations M-ribbons from their shapes in 11 dimensions.

Moreover we generalize the M-ribbons so as to be consistent
with the $SL(2,\mathbb{Z})$ modular transformation on $T^2$. After T-duality
transformation in the 9th direction, these configurations are interpreted
as $(p,q)$-strings which wind and move at light velocity in the 9th direction
with an arbitrary profile in type IIB theory.
These configurations can be interpreted as an
$SL(2,\mathbb{Z})$ duality family of the super D-helix.

The M-ribbons constructed in this paper are configurations which are
straight in one direction. It is an interesting question
whether it is possible to deform the configuration in that direction.
Here we also examine the problem and show that such deformation
does not give BPS solutions.

\section{Supertubes and Supercurves via M-Ribbon}

We start with the construction of classical configurations of an M2-brane,
which would correspond to supertubes and supercurves in type IIA theory.

Let us consider M-theory on $\mathbb{R}^{1,8} \times T^2$ with radii $R_9$
and $R_{11}$, and choose the space-time coordinates such that the line
element becomes
\bea
ds^2 = - dt^2 + d\vec{x} \cdot d\vec{x}
  + R_9^2 d\phi_9^2 + R_{11}^2d\phi_{11}^2,
\label{le}
\ena
where $\vec{x} \in \mathbb{R}^8$ and $0 \leq \phi_9, \phi_{11} \leq 2\pi$.
The radius of the 11th direction is related to the parameters of type IIA
theory, string length $\ell_s$ and string coupling constant $g_s$,
as $R_{11} = g_s \ell_s$. The usual uncompactified case can be recovered by
sending $R_9 \to \infty$.

The bosonic part of M2-brane action is described by the sum of the
Nambu-Goto and Wess-Zumino terms. Throughout this paper,
the relevant part to our analyses is only the Nambu-Goto term:
\begin{alignat}{3}
S = - T_2 \int d^3\xi \sqrt{-\det P[G]_{ab}},
  \label{eq:actM2}
\end{alignat}
where $\xi^a (a,b \!=\! 0,1,2)$ are the world-volume coordinates of M2-brane.
Especially we take the parameter region for $\xi^1$ and $\xi^2$ from $0$ to
$2\pi$. $P[G]_{ab}$ denotes the induced metric on the M2-brane, and
$T_2$ is the tension of M2-brane given as $T_2 = 1/(2\pi)^2 \ell_p^3
= 1/(2\pi)^2 \ell_s^3 g_s$ in terms of the 11-dimensional Planck length
$\ell_p$ and parameters of type IIA theory.

Now we consider the M2-brane embedded in the target space as follows:
\begin{alignat}{3}
t = \xi^0, \quad
\vec{x} = \vec{x}(\xi^1) ,\quad \phi_9 = \xi^2,  \quad
\phi_{11} = \phi_{11}(\xi^0,\xi^1).
\label{an1}
\end{alignat}
This assumption means that M2-brane is straight in the 9th direction and
spreads in an arbitrary curve in $\mathbb{R}^8$, if the 11th direction
is neglected. We also impose the condition
\begin{alignat}{3}
\phi_{11}(\xi^0,\xi^1 \!+\! 2\pi) = \phi_{11}(\xi^0,\xi^1) + 2\pi q,
\label{eq:cond}
\end{alignat}
where the integer $q$ is the winding number around the 11th direction.
The induced metric on the M2-brane becomes
\begin{alignat}{3}
P[G]_{ab}d\xi^a d\xi^b =
  \begin{pmatrix}
    d\xi^0 \!\!&\!\! d\xi^1 \!\!&\!\! d\xi^2
  \end{pmatrix}
  \!\!
  \begin{pmatrix}
    -1 \!+\! R_{11}^2 \dot{\phi}_{11}^2 &
    R_{11}^2 \dot{\phi}_{11} {\phi'}_{\!\!11} & 0
    \\
    R_{11}^2 \dot{\phi}_{11} {\phi'}_{\!\!11} &
    {|\vec{x}'|}^2 \!+\! R_{11}^2 {\phi'}_{\!\!11}^2 & 0
    \\
    0 & 0 & R_9^2
  \end{pmatrix}
  \!\!\!
  \begin{pmatrix}
    d\xi^0 \\ d\xi^1 \\ d\xi^2
  \end{pmatrix},
\end{alignat}
and the M2-brane action can be evaluated to be
\begin{alignat}{3}
S &= -T_2 \int d^3\xi \sqrt{X}, \notag
  \\
  X &\equiv -\det P[G]_{ab} = R_9^2 {|\vec{x}\,'|}^2 \!+\! R_9^2 R_{11}^2
  {\phi'}_{\!\!11}^2 - R_9^2 R_{11}^2 {|\vec{x}\,'|}^2 {\dot \phi}_{11}^2,
\label{eq:act}
\end{alignat}
where partial derivatives $\partial_{\xi^0}$ and $\partial_{\xi^1}$
are denoted as $\,\dot{}$ and $\,'$, respectively.

Before solving the equations of motion obtained from the above action,
let us examine the energy of M2-brane in the Hamiltonian formalism.
The momentum conjugate to $\phi_{11}$ is
\begin{alignat}{3}
p_{11} \equiv \frac{\delta L}{\delta {\dot \phi}_{11}} = T_2
  \frac{R_9^2 R_{11}^2 {|\vec{x}\,'|}^2 {\dot \phi}_{11}}{\sqrt{X}} ,
  \label{eq:p11}
\end{alignat}
and we obtain the Hamiltonian:
\bea
H &=& \int d\xi^1 d\xi^2 \; T_2
\frac{R_9^2 {|\vec{x}\,'|}^2 + R_9^2 R_{11}^2 {\phi'}_{\!\!11}^2}{\sqrt{X}} \nn
&=& \int d\xi^1 d\xi^2
  \sqrt{ \bigg(\frac{1}{R_{11}^2} + \frac{{\phi'}_{\!\!11}^2}
  {{|\vec{x}\,'|}^2} \bigg) p_{11}^2
  + T_2^2 R_9^2 \big( {|\vec{x}\,'|}^2 + R_{11}^2
  {\phi'}_{\!\!11}^2 \big) } .
\label{ham1}
\ena
The integrand represents the energy of M2-brane per unit area, which is given
by the square root of the sum of momentum squared and mass squared.
If $p_{11} = 0$, the energy of M2-brane attains its minimum when
$|\vec{x}\,'| \!=\! 0$, that is, M2-brane shrinks to zero size
in $\mathbb{R}^8$ and becomes a fundamental string in type IIA perspective.
Also if $\phi'_{11} = 0$, the energy of M2-brane takes its minimum for
$|\vec{x}\,'| \!=\! 0$, that is, M2-brane again shrinks to zero size
in $\mathbb{R}^8$ and becomes D0-branes in type IIA theory.
If $\phi'_{11} p_{11} \neq 0$, however, the momentum term prevents
$|\vec{x}\,'|$ from being zero.\footnote{In type IIA theory,
$\phi'_{11}$ and $p_{11}$ correspond to the electric and magnetic fluxes
on D2-brane, respectively. It is interesting to note that
by comparing the action (\ref{eq:act}) and that of supertube,
which is obtained by replacing $X$ in (\ref{eq:act}) as
$X = R_9^2|\vec x'{}^2| + (2\pi \ell_s^2 F_{12})^2
- |\vec x'{}^2|(2\pi \ell_s^2 F_{02})^2$, $\dot{\phi}_{11}$
and $\phi'_{11}$ play the roles of the electric field $F_{02}$ and
magnetic field $F_{12}$, respectively.}
We are now going to show that in fact
there exists expanded BPS M-ribbon with an arbitrary shape in $\mathbb{R}^8$.

Let us first note that the Hamiltonian~\p{ham1} can be rewritten as
\begin{alignat}{3}
H &= \int d\xi^1 d\xi^2
  \sqrt{ \bigg( \frac{\phi'_{11}}{|\vec{x}\, '|} p_{11}
  - T_2 R_9 {|\vec{x}\, '|} \bigg)^2 + \bigg(\frac{p_{11}}{R_{11}}
  + T_2 R_9 R_{11} \phi'_{11} \bigg)^2 } \notag
  \\
  &\geq \int d\xi^1 d\xi^2
  \Big( \frac{p_{11}}{R_{11}} + T_2
  R_9 R_{11} \phi'_{11} \Big).
\end{alignat}
Here for simplicity we have assumed $p_{11} > 0$ and $\phi'_{11} > 0$,
but we can also consider other cases similarly.
The inequality in the above equation is saturated for
\bea
p_{11} = T_2 \frac{R_9 |\vec{x}\, '|^2}{\phi'_{11}}. \label{eq:rad}
\ena
Together with eqs.~\p{eq:act} and \p{eq:p11}, we find
\bea
\dot \phi_{11} = \frac{1}{R_{11}},\quad
  \sqrt{X} = R_9 R_{11} \phi'_{11}.
\label{eq:tube}
\ena
The former equation shows that the M-ribbon is moving in the 11th direction
at the speed of light, as can be seen from the line element \p{le}.
With the condition (\ref{eq:cond}), the energy for this configuration can
be written as
\bea
H = \frac{N}{R_{11}} + 2\pi R_9 Tq,
\label{eq:enetube}
\ena
where $T = 1/2\pi \ell_s^2$ is the string tension and
$N = \int d\xi^1 d\xi^2 p_{11}$ represents the total 11th momentum.
It should be noticed that the energy (\ref{eq:enetube}) is derived by
assuming $p_{11} > 0$ and ${\phi'}_{\!\! 11} > 0$, i.e., $N > 0$ and $q > 0$.
Energy for general cases is obtained by replacing $N$ and $q$ with their
absolute values. Thus we see that the energy~\p{eq:enetube} is just the sum of
the masses of $|N|$ D0-branes and $|q|$ strings, which matches exactly
that of supertubes. As a further confirmation, let us consider the
M2-brane which is embeded in $\mathbb{R}^8$ with circular cross-section.
{}From eq. (\ref{eq:rad}), we see that the radius of M2-brane is given by
\begin{alignat}{3}
|\vec{x}'| = \frac{1}{T_2} \sqrt{q_0 q_s}, \quad
  q_0 \equiv \frac{N}{2\pi R_9 R_{11}},\quad q_s \equiv Tq,
\end{alignat}
where $q_0$ is the D0-brane charge per unit length and
$q_s$ is the string charge, respectively.
This correctly reproduces the radius of supertube with circular
cross-section \cite{MT}.

\begin{figure}[bt]
\begin{center}
\includegraphics[width=13cm,height=6cm,keepaspectratio]{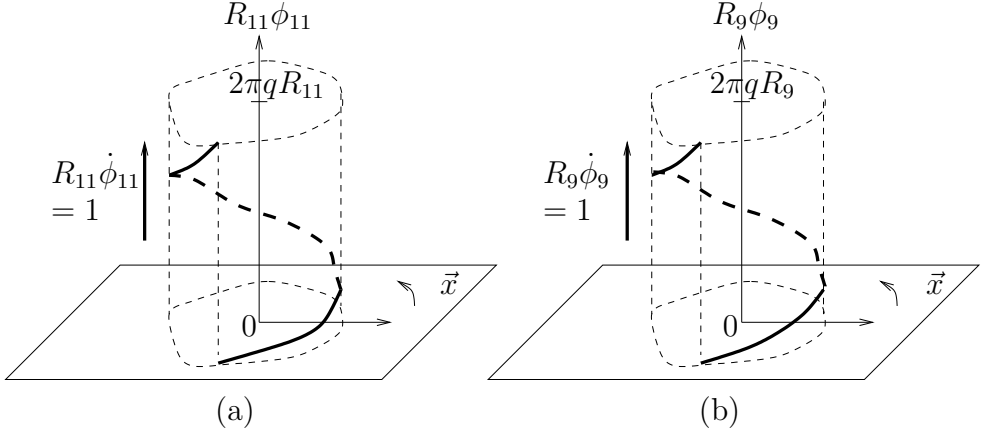}
\begin{picture}(400,0)
    \put(95,0){(a)}
    \put(278,0){(b)}
    \put(180,48){$\vec{x}$}
    \put(98,150){$R_{11}\phi_{11}$}
    \put(105,32){$0$}
    \put(100,124){$2\pi q R_{11}$}
    \put(33,90){$R_{11}{\dot{\phi}}_{11}$}
    \put(33,76){$=1$}
    \put(364,48){$\vec{x}$}
    \put(283,150){$R_9 \phi_9$}
    \put(287,32){$0$}
    \put(282,124){$2\pi q R_9$}
    \put(219,90){$R_9{\dot{\phi}}_9$}
    \put(219,76){$=1$}
\end{picture}
\caption{(a) M-ribbon wrapped around the 11th dimension, is equivalent to the
supertube when the 11th dimension is reduced, and (b) to supercurve.
They transform into each other under 9-11 flip.}
\label{fig:fig1}
\end{center}
\end{figure}

Now let us show that the M2-brane configuration with an arbitary cross-section
solves the field equations following from the action~\p{eq:act}:
\begin{alignat}{3}
&\quad\; \pa_{\xi^1} \bigg( \frac{(1- R_{11}^2\dot{\phi}_{11}^2) \vec{x}\, '}
  {\sqrt{X}}\bigg) = 0, \notag  \\
&- \partial_{\xi^0} \bigg( \frac{R_{11}^2 |\vec{x}\, '|^2 \dot{\phi}_{11}}
  {\sqrt{X}}\bigg) + \partial_{\xi^1} \bigg( \frac{R_{11}^2 \phi'_{11}}
  {\sqrt{X}}\bigg) = 0.
\end{alignat}
It is easy to see that both of these equations are satisfied by \p{eq:tube}
for an arbitrary function $\vec{x}=\vec{x}(\xi^1)$ as long as $\phi'_{11}$
does not depend on $\xi^0$. This configuration is depicted in
Fig.~\ref{fig:fig1}(a).

When the 11th direction is reduced, this ribbon-shaped configuration produces
a closed curve in $\mathbb{R}^8$ which is extended in the 9th direction
$R_9 \phi_9$. This M-ribbon can thus be interpreted as a D2-brane which is
straight in the 9th direction and spreads in an arbitrary curve in
$\mathbb{R}^8$. Furthermore, this D2-brane carries charges of D0-branes and
fundamental strings along the 9th direction, as can be seen from
eq.~(\ref{eq:enetube}). We thus conclude that we have reproduced supertubes
with an arbitrary cross-section via M-ribbon.\footnote{Supertubes are
constructed originally in the flat space $\mathbb{R}^{1,9}$ in type IIA
theory. Our configurations reproduce those in the limit $R_9 \to \infty$.}

On the other hand, the above discussions can be repeated with the roles of
9th and 11th directions interchanged (see Fig.~\ref{fig:fig1}(b)).
Then the Hamiltonian is given by
\bea
H = \frac{N}{R_9} + 2\pi R_9 Tq.
\label{eq:enecurve}
\ena
Upon reducing the 11th direction, the M-ribbon configuration gives a wrapped
curve, and is interpreted as a fundamental string
extending arbitrarily in $S \times \mathbb{R}^8$ and wound $q$ times around
the circle in the 9th direction. Furthermore, this fundamental string carries
momentum in the 9th direction, as can be seen from eq.~(\ref{eq:enecurve}).
This energy matches precisely with that of supercurve, which is originally
obtained by making T-S-T-duality transformation to the supertube~\cite{MNT2}.
In fact, what we have produced here is equivalent to the supercurve
because such duality map corresponds to 9-11 flip in the context of M-theory.

In this way, our configuration in M-theory gives a unified picture of
the supertubes and supercurves.

\section{$SL(2,\mathbb{Z})$ Family of M-Ribbons}

It is well known that the $SL(2,\mathbb{Z})$ modular
transformation on $T^2$ is exactly related to the $SL(2,\mathbb{Z})$ duality
transformation on type IIB theory, which includes the S-duality.
Therefore it is interesting to generalize the supertube and supercurve
configurations in a way consistent with that duality.

Our line element of the flat space-time is again taken as eq.~(\ref{le}),
but the M2-brane is embedded into the target space as
\begin{alignat}{3}
  &t = \xi^0, \quad \vec{x} = \vec{x}(\xi^1), \quad
  \phi_9 = \varphi_9(\xi^0,\xi^1) + m\xi^2, \quad
  \phi_{11} = \varphi_{11}(\xi^0,\xi^1) + n\xi^2,
\end{alignat}
with the conditions
\begin{alignat}{3}
  &\varphi_9(\xi^0,\xi^1 \!+\! 2\pi) = \varphi_9(\xi^0,\xi^1) + 2\pi p,\quad\,\,
  \varphi_{11}(\xi^0,\xi^1 \!+\! 2\pi) = \varphi_{11}(\xi^0,\xi^1) + 2\pi q.
  \label{eq:cond2}
\end{alignat}
When $\xi^1$ goes from $0$ to $2\pi$, one direction of the M2-brane makes
an arbitrary curve in $\mathbb{R}^8$ and simultaneously winds $p$ times
around the 9th direction and $q$ times around the 11th direction.
Similarly when $\xi^2$ goes from $0$ to $2\pi$, another direction of the
M2-brane winds $m$ times around the 9th direction and $n$ times around the
11th direction. For example, the supertube corresponds to the choice
$(p,m;q,n)=(0,1;q,0)$ and the supercurve to $(p,m;q,n)=(p,0;0,1)$.
The $SL(2,\mathbb{Z})$ matrix acts on $T^2$ as
\begin{alignat}{3}
  \begin{pmatrix}
    \phi_9 \\
    \phi_{11}
  \end{pmatrix}
  \to
  \begin{pmatrix}
    a & b \\
    c & d
  \end{pmatrix}
  \begin{pmatrix}
    \phi_9 \\
    \phi_{11}
  \end{pmatrix},
\end{alignat}
where $ad-bc = 1$ and $mq - np$ is invariant under this transformation.

The induced metric becomes
\begin{alignat}{3}
&P[G]_{ab}d\xi^a\xi^b \notag
  \\
  &= -(d\xi^0)^2 \!+\! |\vec{x}\, '|^2 (d\xi^1)^2 \!+\!
  R_9^2 (\dot{\varphi}_9 d\xi^0 \!+\! {\varphi'}_{\!\! 9} d\xi^1
  \!+\! m d\xi^2)^2 \!+\!
  R_{11}^2 (\dot{\varphi}_{11}d\xi^0 \!+\! {\varphi'}_{\!\! 11} d\xi^1
  \!+\! n d\xi^2)^2
  \\
  &=
  \begin{pmatrix}
    d\xi^0 \!&\!\! d\xi^1 \!&\!\! d\xi^2
  \end{pmatrix} \!\!\!
  \begin{pmatrix}
    -1 \!+\! R_9^2 \dot{\varphi}_9^2 \!+\! R_{11}^2 \dot{\varphi}_{11}^2
    &
    R_9^2 \dot{\varphi}_9 {\varphi'}_{\!\! 9}
    \!+\! R_{11}^2 \dot{\varphi}_{11} {\varphi'}_{\!\! 11}
    &
    m R_9^2 \dot{\varphi}_9 \!+\! n R_{11}^2 \dot{\varphi}_{11}
    \\
    R_9^2 \dot{\varphi}_9 {\varphi'}_{\!\! 9}
    \!+\! R_{11}^2 \dot{\varphi}_{11} {\varphi'}_{\!\! 11}
    &
    |\vec{x}\, '|^2 \!+\! R_9^2 {\varphi'}_{\!\! 9}^2
    \!+\! R_{11}^2 {\varphi'}_{\!\! 11}^2
    &
    m R_9^2 {\varphi'}_{\!\! 9} \!+\! n R_{11}^2 {\varphi'}_{\!\! 11}
    \\
    m R_9^2 \dot{\varphi}_9 \!+\! n R_{11}^2 \dot{\varphi}_{11}
    &
    m R_9^2 {\varphi'}_{\!\! 9} \!+\! n R_{11}^2 {\varphi'}_{\!\! 11}
    &
    m^2 R_9^2 \!+\! n^2 R_{11}^2
  \end{pmatrix} \!\!\!\!
  \begin{pmatrix}
    d\xi^0 \\ d\xi^1 \\ d\xi^2
  \end{pmatrix} \!\!,
  \notag
\end{alignat}
and the volume factor is calculated as
\bea
&&\sqrt{-\det P[G]_{ab}} \nn
&&= \sqrt{(m^2 R_9^2 \!+\! n^2 R_{11}^2)|\vec{x}\, '|^2
  \!+\! R_9^2 R_{11}^2 (n{\varphi'}_{\!\! 9} \!-\! m{\varphi'}_{\!\! 11})^2
  \!-\! R_9^2 R_{11}^2 |\vec{x}\, '|^2
  (n\dot{\varphi}_9 \!-\! m\dot{\varphi}_{11})^2 },
\ena
where $\;\dot{}\,=\partial_{\xi^0}$ and $\;'\,=\partial_{\xi^1}$ as before.
It is useful to introduce new coordinates
\begin{alignat}{3}
  \begin{cases}
    \psi = - n\phi_9 + m\phi_{11} = - n\varphi_9(\xi^0,\xi^1)
    + m\varphi_{11}(\xi^0,\xi^1),
    \\[0.3cm]
    \chi = \dfrac{m R_9^2 \phi_9 \!+\! n R_{11}^2 \phi_{11}}
    {\sqrt{m^2 R_9^2 \!+\! n^2 R_{11}^2}}
    = \dfrac{m R_9^2 \varphi_9(\xi^0,\xi^1) \!+\! n R_{11}^2 \varphi_{11}(\xi^0,\xi^1)}
    {\sqrt{m^2 R_9^2 \!+\! n^2 R_{11}^2}}
    + \sqrt{m^2 R_9^2 \!+\! n^2 R_{11}^2}\xi^2 .
  \end{cases}
\end{alignat}
Note that $\psi$ does not depend on $\xi^2$ and
$\partial_{\xi^2} \chi = \sqrt{m^2 R_9^2 \!+\! n^2 R_{11}^2}$.
The relevant part of the line element is rewritten as
\begin{alignat}{3}
R_9^2 d\phi_9^2 + R_{11}^2 d\phi_{11}^2 =
  R_\psi^2 d\psi^2 + d\chi^2, \label{eq:met}
\end{alignat}
where $R_\psi$, which represents the radius of the $\psi$ direction,
is defined by
\begin{alignat}{3}
  R_\psi \equiv \frac{R_9 R_{11}}{\sqrt{m^2 R_9^2 + n^2 R_{11}^2}}.
\end{alignat}
Then the M2-brane action is given by
\begin{alignat}{3}
S &= -T_2 \int d^3\xi \sqrt{X}, \notag
  \\
  X &\equiv (m^2 R_9^2 \!+\! n^2 R_{11}^2)|\vec{x}\, '|^2
  \!+\! R_9^2 R_{11}^2 {\psi'}^2  \!-\! R_9^2 R_{11}^2 |\vec{x}\, '|^2
  {\dot{\psi}}^2 .\label{eq:actM}
\end{alignat}

We use the Hamiltonian formalism to estimate the energy of M2-brane.
The momentum conjugate to $\psi$ is defined as
\begin{alignat}{3}
p_{\psi} &\equiv \frac{\delta L}{\delta \dot{\psi}} =
  T_2 \frac{R_9^2 R_{11}^2 |\vec{x}\, '|^2 \dot{\psi}}{\sqrt{X}},
\label{eq:ppsi}
\end{alignat}
and the Hamiltonian takes the form
\bea
H 
= \int\!\! d\xi^1 d\xi^2 \sqrt{
  \bigg(\! \frac{1}{R_\psi^2}
  \!+\! \frac{{\psi'}^2}{|\vec{x}\, '|^2} \!\bigg) p_\psi^2
  \!+\! T_2^2 (m^2 R_9^2 \!+\! n^2 R_{11}^2) \big( |\vec{x}\, '|^2
  \!+\! R_\psi^2 {\psi'}^2 \big) }.
\ena
The integrand represents the energy of M2-brane per unit area, which is given
by the square root of the sum of momentum in the $\psi$ direction squared
and mass squared. If $\psi' p_{\psi} = 0$, the energy of M2-brane becomes
minimal when $|\vec{x}\, '| = 0$, that is, M2-brane collapses to zero size in
$\mathbb{R}^8$. If $\psi' p_{\psi} \neq 0$, however, the M2-brane will be
stable against collapse with some shape in $\mathbb{R}^8$.

To see this, we rewrite the Hamiltonian as
\begin{alignat}{3}
H &= \int d\xi^1 d\xi^2
  \sqrt{ \bigg( \frac{\psi'}{|\vec{x}\, '|} p_{\psi}
  \!-\! T_2 \sqrt{m^2 R_9^2 \!+\! n^2 R_{11}^2}
  {|\vec{x}\, '|} \bigg)^2 + \bigg(\frac{p_{\psi}}{R_{\psi}}
 \!+\! T_2 \sqrt{m^2 R_9^2 \!+\! n^2 R_{11}^2} R_{\psi} \psi' \bigg)^2 } \notag
  \\
  &\geq \int d\xi^1 d\xi^2
  \Big( \frac{p_{\psi}}{R_{\psi}}
  \!+\! T_2 \sqrt{m^2 R_9^2 \!+\! n^2 R_{11}^2} R_{\psi} \psi' \Big).
\end{alignat}
The inequality of the above equation is saturated for
\bea
p_{\psi}= T_2 \frac{\sqrt{m^2 R_9^2 \!+\! n^2 R_{11}^2}|\vec{x}\, '|^2}{\psi'}.
\ena
Together with eqs.~\p{eq:actM} and \p{eq:ppsi}, we find
\bea
\dot \psi = \frac{1}{R_{\psi}},\quad
  \sqrt{X} = R_9 R_{11} \psi'.
  \label{eq:Mtube}
\ena
The former equation shows that the M2-brane is moving in the $\psi$ direction
at the speed of light. With the help of the condition (\ref{eq:cond2}),
the energy for this configuration can be written as
\bea
H = \frac{N}{R_{\psi}} + (mq \!-\! np)T 2\pi R_9, \label{eq:Menetube}
\ena
where $N = \int d\xi^1 d\xi^2 p_{\psi}$ represents the total momentum
in the $\psi$ direction. The energy~(\ref{eq:Menetube}) is for
$p_{\psi} > 0$ and ${\psi'} > 0$, i.e., $N > 0$ and $(mq - np) > 0$, but
for general cases it is given by replacing $N$ and $(mq - np)$ with their
absolute values.

It is also easy to see that the field equations following from
the action~\p{eq:actM}:
\bea
\quad\; \partial_{\xi^1} \bigg( \frac{(1-R_\psi^2 \dot{\psi}^2) \vec{x}\,'}
  {\sqrt{X}} \bigg) = 0, \label{eq:eom2} \nn
- \partial_{\xi^0} \bigg( \frac{|\vec{x}\,'|^2 \dot{\psi}}
  {\sqrt{X}}\bigg) +
  \partial_{\xi^1} \bigg( \frac{{\psi'}}
  {\sqrt{X}}\bigg) = 0, \label{eq:eom}
\ena
are satisfied by \p{eq:Mtube} for an arbitrary function
$\vec{x}=\vec{x}(\xi^1)$ as long as $\psi'$ does not depend on $\xi^0$.

In this way we have generalized the configurations of M-ribbons to be
consistent with $SL(2,\mathbb{Z})$ duality.
This fact can be seen more easily in the context of type IIB theory.
By using the T-duality relations,
\bea
  R_9 = \frac{\ell_s^2}{R_9'}, \quad g_s = \frac{\ell_s}{R_9'}g_s',
\ena
the energy (\ref{eq:Menetube}) can be written as
\begin{alignat}{3}
H = N\sqrt{(nT2\pi R_9')^2 + (mT_1 2\pi R_9')^2} + \frac{mq \!-\! np}{R_9'},
\end{alignat}
where $g_s'$ and $R_9'$ are the string coupling constant and the radius of
the 9th direction in type IIB theory, and $T_1$ is the tension of D-string.
This energy precisely agrees with that of a configuration of $(n,m)$-string,
which is wrapped $N$ times around 9th direction with an arbitrary profile
and moves in that direction with $(mq - np)$ units of momentum.
The super D-helix, which is a helical D-string with momentum along
the 9th direction, is realized for $N \!=\! 1$ and $(p,m;q,n)\!=\! (0,1;q,0)$.
These numbers are the same as those of supertube and it is
consistent with the fact that the supertube and the super D-helix transform
into each other simply under the T-duality.

\section{Quarter Supersymmetry}

Let us check whether the configurations considered so far are
supersymmetric or not. The background is the 11-dim. flat space-time,
so there are 32 supercharges. We denote the constant Majorana spinor
which parametrizes the supersymmetric transformation as $\e$.
The supersymmetric transformation of fermion $\Theta$ on the M2-brane
is given by
\begin{alignat}{3}
  \delta \Theta = \epsilon + (1+\Gamma) \kappa,
\end{alignat}
where $\Gamma$ is defined by
\begin{alignat}{3}
  \Gamma = \frac{1}{3! \sqrt{-\det P[G]_{ab}}} \epsilon^{abc}
  \partial_a X^L \partial_b X^M \partial_c X^N \Gamma_{LMN},
\end{alignat}
and satisfies the relation $\Gamma^2 = 1$.
We then see that the condition $\delta \Theta =0$ is written as
\begin{alignat}{3}
  (1-\Gamma) \epsilon = 0. \label{eq:susy}
\end{alignat}
Noting eq.~(\ref{eq:met}), we obtain
\begin{alignat}{3}
  \Gamma &= \frac{1}{\sqrt{X}}
  \Big( \psi' \partial_{\xi^2}\chi \Gamma_{t\psi\chi} \!+\!
  \partial_{\xi^2}\chi\, \vec{x}\,' \cdot \Gamma_{t \vec{x} \chi} \!+\!
  \dot{\psi} \partial_{\xi^2}\chi\, \vec{x}\,' \cdot
  \Gamma_{\psi \vec{x} \chi} \Big) \notag
  \\
  &= \frac{R_9 R_{11} \psi'}{\sqrt{X}} \Gamma_{\hat{t} \hat{\psi} \hat{\chi}}
  \!+\! \frac{R_9 R_{11}}{R_\psi \sqrt{X}}
  \vec{x}\,' \cdot \Gamma_{\hat{t} \hat{\vec{x}} \hat{\chi}}
  \big( 1 \!-\! R_\psi \dot{\psi} \Gamma_{\hat{t} \hat{\psi}} \big),
\end{alignat}
where $X$ is given by (\ref{eq:actM}) and
$\vec{x}\,' \cdot \Gamma_{\vec{x}} \!=\! \sum_{i=1}^8 (x^i)'
\Gamma_{x^i}$. Here we introduced $\Gamma_{\hat{M}}$ which
satisfy $SO(1,10)$ Clifford algebra and represented by constant matrices.
$\Gamma_M$ and $\Gamma_{\hat{M}}$ are equivalent but
$\Gamma_\psi = R_\psi \Gamma_{\hat{\psi}}$.
Then eq.~(\ref{eq:susy}) is cast into
\begin{alignat}{3}
  \Big(1 - \frac{R_9 R_{11} \psi'}{\sqrt{X}}
  \Gamma_{{\hat t}{\hat{\psi}}{\hat \chi}}\Big)\epsilon
  - \frac{R_9 R_{11}}{R_\psi \sqrt{X}}
  \vec{x}\,' \cdot \Gamma_{\hat{t} \hat{\vec{x}} \hat{\chi}}
  \big( 1 \!-\! R_\psi \dot{\psi} \Gamma_{\hat{t} \hat{\psi}} \big)
  \epsilon = 0.
\end{alignat}
>From this, we see that the BPS equation has nontrivial solution
when the relation $R_\psi \dot{\psi} = 1$ holds, and
then $\epsilon$ is explicitly written as
\begin{alignat}{3}
  \epsilon = \frac{1 + \Gamma_{{\hat t}{\hat{\psi}}{\hat \chi}}}{2}
  \frac{1 + \Gamma_{\hat{t}\hat{\psi}}}{2} \epsilon_0,
\end{alignat}
for an arbitrary constant Majorana spinor $\epsilon_0$.
As explained in the previous section, the relation $R_\psi \dot{\psi} = 1$
is the solution of field equations (\ref{eq:eom}).

Thus all configurations of M-ribbons preserve $\frac{1}{4}$ supersymmetry.
Our results generalize those for supertubes and
supercurves~\cite{MT,BL,EMT,MNT1,MNT2}.

\section{Discussions}

In this paper, we have obtained the $\frac14$ BPS configurations of M-ribbons
on $\mathbb{R}^{1,8} \times T^2$, which reduce to the supertubes and
supercurves in type IIA theory upon dimensional reduction.
In the context of M-theory, the supertubes are related to the
supercurve under 9-11 flip.
Furthermore, those configurations are generalized so as to be
consistent with the $SL(2,\mathbb{Z})$ modular transformation on $T^2$.
Thus our configurations give a unified picture of all these configurations.

Let us consider if our M-ribbon configurations~(\ref{an1}) can be
generalized to have dependence on $\xi^2$. To be explicit, consider
$\vec{x} = \vec{x}(\xi^1,\xi^2)$, giving the configurations of supertubes
whose cross-sections depend on $\xi^2$. After straightforward calculation
similar to section 2, we obtain the following BPS bound on the Hamiltonian:
{\small
\begin{alignat}{3}
H &\!=\!\! \int \!\! d\xi^1 \! d\xi^2
  \sqrt{ \bigg(\frac{1}{R_{11}^2} \!+\!
  \frac{ (|\vec b|^2 \!+\! R_9^2) {\phi'}_{\!\!11}^2}
  { |\vec a|^2 (|\vec b|^2 \!+\! R_9^2) \!-\! |a b|^2}
  \bigg) p_{11}^2
  \!+\! T_2^2 \bigg(
  |\vec a|^2 (|\vec b|^2 \!+\! R_9^2) \!-\! |a b|^2
  \!+\! (|\vec b|^2 \!+\! R_9^2)R_{11}^2 {\phi'}_{\!\!11}^2
  \bigg) } \notag
  \\
  &\!=\!\! \int \!\! d\xi^1 \! d\xi^2
  \sqrt{ \bigg( \frac{\sqrt{|\vec b|^2 \!+\! R_9^2}\phi'_{11}p_{11}}
  {\sqrt{|\vec a|^2 (|\vec b|^2 \!+\! R_9^2) \!-\! |a b|^2}}
  \!-\! T_2 \sqrt{|\vec a|^2 (|\vec b|^2 \!+\! R_9^2)
  \!-\! |a b|^2} \bigg)^2 \!\!+\! \bigg(\frac{p_{11}}{R_{11}}
  \!+\! T_2 \sqrt{|\vec b|^2 \!+\! R_9^2} R_{11} \phi'_{11} \bigg)^2 } \notag
  \\
  &\!\geq\!\! \int \!\! d\xi^1 \! d\xi^2
  \Big( \frac{p_{11}}{R_{11}} + T_2 \sqrt{|\vec b|^2 \!+\! R_9^2}
  R_{11} \phi'_{11} \Big),
\end{alignat}
}
where $\vec a \!=\! \pa_{\xi^1}\vec{x}$, $\vec b \!=\! \pa_{\xi^2} \vec{x}$
and $ab \!=\! \partial_{\xi^1}\vec{x} \cdot \partial_{\xi^2} \vec{x}$.
The inequality is saturated for
\bea
\dot{\phi}_{11} = \frac{1}{R_{11}}.
\label{fcond}
\ena
We can check that the field equations and the $\frac14$ BPS (Killing spinor)
equation are also satisfied only for \p{fcond} and constant $\vec b$.
This means that the dependence of $\vec x$ on $\xi^2$ is at most linear and
the embedding of M2-brane into the target space should be straight line
in $\xi^2$. We can even erase the apparent dependence by appropriate
rotation of the coordinates.
We thus conclude that the M-ribbons whose cross-sections depend
on the extending direction are not BPS solutions.

We can also draw the same conclusion from the viewpoint of Matrix theory.
Let us consider tubular configuration in $(x,y,z)$ space, which
extends in the $z$ direction.
In the context of Matrix theory, one of the BPS conditions is given by
\begin{alignat}{3}
[X,Y] = 0,
\end{alignat}
where $X$ and $Y$ are collective coordinates of D0-branes in $(x,y)$-plane.
In ref.~\cite{Hya2}, various configurations of tubular D2-branes, whose
cross-sections are dependent on $z$, are considered. There, by comparing with
the analytical results for D2-brane, it was numerically shown that $[X,Y]$
can be zero only when the radius of tubular configuration is constant,
consistent with our above result.

Recently it has been argued that the supertubes can end on a D4-brane
\cite{KMPW}. It is interesting to understand these situations by lifting to
M-theory\cite{AH}. It is also interesting to reconstruct the configurations
of BPS M-ribbons from the viewpoint of Matrix theory.

\section*{Acknowledgements}

YH would like to thank S. Sugimoto, K. Hosomichi, S. Ogushi and
M. Ninomiya for discussions and encouragements.
NO thanks D. Bak for discussions.
The work of NO was supported in part by a Grant-in-Aid for Scientific
Research No. 12640270.

\newcommand{\NP}[1]{Nucl.\ Phys.\ {\bf #1}}
\newcommand{\PL}[1]{Phys.\ Lett.\ {\bf #1}}
\newcommand{\CQG}[1]{Class.\ Quant.\ Grav.\ {\bf #1}}
\newcommand{\JHEP}[1]{JHEP\ {\bf #1}}
\newcommand{\PR}[1]{Phys.\ Rev.\ {\bf #1}}
\newcommand{\PRL}[1]{Phys.\ Rev.\ Lett.\ {\bf #1}}

\end{document}